\documentclass[preprint]{aastex}

\begin{document}

\title {The Eclipsing System V404 Lyr: Light-Travel Times and $\gamma$ Doradus Pulsations }
\author{Jae Woo Lee$^{1,2}$, Seung-Lee Kim$^{1,2}$, Kyeongsoo Hong$^1$, Chung-Uk Lee$^{1,2}$, and Jae-Rim Koo$^1$}
\affil{$^1$Korea Astronomy and Space Science Institute, Daejeon 305-348, Korea}
\affil{$^2$Astronomy and Space Science Major, Korea University of Science and Technology, Daejeon 305-350, Korea}
\email{jwlee@kasi.re.kr, slkim@kasi.re.kr, kshong@kasi.re.kr, leecu@kasi.re.kr, koojr@kasi.re.kr}

\begin{abstract}
We present the physical properties of V404 Lyr exhibiting eclipse timing variations and multiperiodic pulsations from 
all historical data including the ${\it Kepler}$ and SuperWASP observations. Detailed analyses of 2,922 minimum epochs showed 
that the orbital period has varied through a combination with an upward-opening parabola and two sinusoidal variations, 
with periods of $P_3$=649 d and $P_4$=2,154 d and semi-amplitudes of $K_3$=193 s and $K_4$=49 s, respectively. 
The secular period increase with a rate of $+$1.41 $\times 10^{-7}$ d yr$^{-1}$ could be interpreted as a combination of 
the secondary to primary mass transfer and angular momentum loss. The most reasonable explanation for both sinusoids is a pair 
of light-travel-time effects due to two circumbinary objects with projected masses of $M_3$=0.47 M$_\odot$ and 
$M_4$=0.047 M$_\odot$. The third-body parameters are consistent with those calculated using the Wilson-Devinney binary code. 
For the orbital inclinations $i_4 \ga$ 43$^\circ$, the fourth component has a mass within the hydrogen-burning limit of 
$\sim$0.07 M$_\odot$, which implies that it is a brown dwarf. A satisfactory model for the {\it Kepler} light curves was obtained 
through applying a cool spot to the secondary component. The results demonstrate that the close eclipsing pair is in a semi-detached, 
but near-contact, configuration; the primary fills approximately 93\% of its limiting lobe and is larger than the lobe-filling 
secondary. Multiple frequency analyses were applied to the light residuals after subtracting the synthetic eclipsing curve from 
the {\it Kepler} data. This revealed that the primary component of V404 Lyr is a $\gamma$ Dor type pulsating star, exhibiting 
seven pulsation frequencies in the range of 1.85$-$2.11 d$^{-1}$ with amplitudes of 1.38$-$5.72 mmag and pulsation constants 
of 0.24$-$0.27 d. The seven frequencies were clearly identified as high-order low-degree gravity-mode oscillations which might 
be excited through tidal interaction. Only eight eclipsing binaries have been known to contain $\gamma$ Dor pulsating components 
and, therefore, V404 Lyr will be an important test-bed for investigating these rare and interesting objects.
\end{abstract}

{}

\section{INTRODUCTION}

Orbital periods of eclipsing binaries (EBs) are one of the most accurately measured physical parameters that are required 
in order to understand the dynamics of the systems and their stellar structures and evolutions. The period variability may be 
represented using an eclipse timing diagram, which displays the differences between the observed and calculated minimum epochs
{\it versus} time. The orbital periods of many EBs have varied in some combination of secular and sinusoidal variations rather 
than in a monotonic manner (Kreiner et al. 2001; Lee et al. 2009b, 2013a). The presence of a third body orbiting an EB causes 
a periodic variation of the eclipsing period due to the increasing and decreasing light-travel times (LTT) to the observer 
(Irwin 1952, 1959). If the historical database for light curves and eclipse times is large and sufficiently long to understand 
the binary's period change, circumbinary objects that are gravitationally bound to EBs can be detected. 
The circumbinary companions may have played an important function in the formation and evolution of close EBs, which would cause 
them to have short initial orbital periods and evolve into their present configurations through angular momentum loss due to 
magnetic braking (Pribulla \& Rucinski 2006).

EBs can precisely and directly measure the masses and radii for each component from time-series photometry and spectroscopy, 
while pulsating stars provide significant information about the interior structure of the stars through their pulsation features. 
Some EBs contain $\delta$ Sct or $\gamma$ Dor type pulsating components with multiple periods, which are very useful 
for asteroseismical studies. The binary stars that exhibit both eclipses and pulsations are very promising targets that could 
lead to unique constraints on theoretical models, and they have increased significantly in number by ground-based survey projects 
(e.g., Kim et al. 2003) and space missions such as {\it Kepler} and {\it CoRot} (e.g., Southworth et al. 2011; Maceroni et al. 2013). 
Most of them have been found to be $\delta$ Sct-type members of Algol-type semi-detached EBs designated by the so-called oEA star 
(Mkrtichian et al. 2004). In contrast, there have only been eight EBs reported with $\gamma$ Dor pulsating components 
(Zhang et al. 2013; Maceroni et al. 2014). In general, $\gamma$ Dor pulsators are main-sequence or sub-giant stars with 
spectral types ranging from A7 to F5 near the red edge of the $\delta$ Sct instability strip in the H-R diagram. 
Their observational properties are very similar to those of $\delta$ Sct stars, except the significant difference in 
pulsation periods and pulsation constants (Handler \& Shobbrook 2002). Recently, stars with both $\delta$ Sct and $\gamma$ Dor 
characteristics have been discovered, three of which exhibit a hybrid nature of those classes in EBs (CoRot 100866999, 
Chapellier \& Mathias 2013; KIC 4544587, Hambleton et al. 2013; KIC 3858884, Maceroni et al. 2014). The possible existence of 
hybrid $\gamma$ Dor/$\delta$ Sct stars is of great interest because they offer additional constraints on stellar structures.

In the field of the old galactic cluster NGC 6791, V404 Lyr (KIC 3228863, TYC 3121-1291-1, 2MASS J19190594+3822005) was 
classified as a $\beta$ Lyr-type EB system with an orbital period of 0.7309 d by Popova \& Kraicheva (1984). However, 
from their incomplete CCD photometry in the $VI$ bandpasses, Csizmadia \& S\'andor (2001) suggested that the binary star is 
not a member of the cluster. Since then, Somosv\'ari \& Csizmadia (2006) made additional observations in order to complete 
the previous light curves but approximately 20 $\%$ of them remained deficient. Through analyzing the $VI$ light curves, 
they presented that V404 Lyr is an Algol-type semi-detached binary with a mass ratio of $q$=0.485, an orbital inclination of 
$i$=82$^\circ$.3, and a temperature difference between the components of $\Delta T$=1,928 K. Recently, the highly precise 
photometric observations of V404 Lyr have been obtained in long cadence mode of the {\it Kepler} satellite, which led to 
a sampling rate of 29.4244 min. Detailed information regarding the {\it Kepler} spacecraft and its performance can be found 
in Borucki et al. (2010) and Koch et al. (2010). The orbital period changes for V404 Lyr were examined by Rappaport et al. (2013) 
and Conroy et al. (2014) using the {\it Kepler} public data. The former found that the orbital period has varied due to 
a periodic oscillation with a cycle length of 668.4 d and showed that its variation would be produced by the existence of 
a physically linked third body with a mass function of $f(M_3)$=0.017 M$_\odot$ and an eccentricity of $e_3$=0.08. The latter 
concluded that the cause of the period change is an LTT effect due to a third body with a period of 644.1 d and a semi-amplitude 
of 195 s in a circular orbit.

In order to obtain a reasonable set of photometric solutions and to examine whether the previous suggestions are appropriate 
for the orbital period change, we have meticulously investigated the long-term photometric behavior of V404 Lyr using 
all historical data. For this, we used 56,333 individual observations (BJD 2,454,953$-$2,456,391) from Quarter 1 through 
Quarter 16 of the {\it Kepler} operations. Our results from eclipse timing variation, light-curve synthesis, and frequency 
analysis for light residuals indicate that V404 Lyr is most likely a multiple system with a $\gamma$ Dor-type pulsating component.

\section{ECLIPSE TIMING VARIATION AND ITS IMPLICATIONS}

From the {\it Kepler} observations of V404 Lyr, we determined 2,882 eclipse timings and their errors using the method presented
by Kwee \& van Woerden (1956). These are given in the first column of Table 1, wherein 12 additional eclipses were obtained using 
the data from the SuperWASP (Wide Angle Search for Planets) public archive (Butters et al. 2010). In addition to these, 
28 CCD timings have been collected from the literature (Agerer \& H\"ubscher 1996, 1997, 1998, 1999, 2000; 
Csizmadia \& S\'andor 2001; Diethelm 2001; Saf\'ar \& Zejda 2002; Borkovits et al. 2003; Locher 2005; Nelson 2005, 2007; 
B\'ir\'o et al. 2006; H\"ubscher et al. 2006; Br\'at et al. 2007). The timings were transformed from HJD based on UTC into 
TDB-based BJD using the online applets\footnote{http://astroutils.astronomy.ohio-state.edu/time/} developed by Eastman et al. (2010). 
For ephemeris computations, weights were calculated as the inverse squares of the timing errors and were then scaled 
from the standard deviations ($\sigma$=0.0010 d) of all timing residuals.

Previous researchers (Rappaport et al. 2013; Conroy et al. 2014) have proposed that the eclipse timing variation of V404 Lyr 
can be represented using an LTT caused by the presence of a third body orbiting the eclipsing pair. First of all, 
we fitted the times of minimum light to the single LTT ephemeris, as follows:
\begin{eqnarray}
C_1 = T_0 + PE + \tau_{3},
\end{eqnarray}
where $\tau_{3}$ is the LTT due to a tertiary companion (Irwin 1952, 1959) and includes five parameters ($a_{12}\sin i_3$, $e$, 
$\omega$, $n$, $T_{\rm peri}$). Here, $a_{12} \sin i_3$, $e$, and $\omega$ are the orbital parameters of the eclipsing pair around 
the mass center of the triple system. The parameters $n$ and $T_{\rm peri}$ denote the Keplerian mean motion of the mass center 
of the binary components and its epoch of periastron passage, respectively. The Levenberg-Marquart (LM) technique (Press et al. 1992) 
was applied to solve the equation for the ephemeris parameters. The result is plotted in Figure 1 and the orbital parameters 
are summarized in the second column of Table 2, together with their related quantities. The absolute dimensions presented in 
the following section have been used for these and subsequent calculations. The residuals from the LTT ephemeris appear as 
$O$--$C_{1,\rm full}$ in the fourth column of Table 1. 

As shown in Figure 1, the single LTT ephemeris could not describe the previously published minima and the timing residuals 
in the lower panel indicate the existence of a (or some) further effect(s). After testing several other forms, such as 
a quadratic term {\it plus} a single-LTT ephemeris and a two-LTT ephemeris, we found that the eclipse timing variation is best 
fitted using a quadratic {\it plus} two-LTT ephemeris (Lee et al. 2009a, 2013a):
\begin{eqnarray}
C_2 = T_0 + PE + AE^2 + \tau_{3} + \tau_{4}.
\end{eqnarray}
The LM method was again applied to evaluate the thirteen parameters of the ephemeris which are listed in the third and fourth columns
of Table 2. This ephemeris resulted in a much smaller $\chi^2_{\rm red}$=0.996 than the single LTT ephemeris ($\chi^2_{\rm red}$=2.256).
The short-term periods ($P_3$) and semi-amplitudes ($K_3$) for the two ephemerides are in excellent agreement with each other. 
The $O$--$C_2$ diagram constructed using the linear terms is presented at the top of Figure 2 with the solid curve due to the sum 
of the non-linear terms and the dashed parabola due to the quadratic term of equation (2). The second and third panels display 
the $\tau_3$ and $\tau_4$ orbits, respectively, and the bottom panel represents the residuals from the full ephemeris. 
These appear as $O$--$C_{\rm 2,full}$ in the fifth column of Table 1. 

Assuming the orbits of the circumbinary objects are coplanar with that ($i$=82$^\circ$.9) of the eclipsing pair of V404 Lyr, 
the masses of the third and fourth bodies are $M_3$=0.469 M$_\odot$ and $M_4$=0.047 M$_\odot$, respectively, and the radii 
are calculated to be $R_3$=0.484 R$_\odot$ and $R_4$=0.054 R$_\odot$ from the empirical relation described by Southworth (2009). 
If the third component is on main sequence, its temperature and bolometric luminosity are $T_3$=3,445 K and $L_3$=0.030 $L_{\sun}$, 
respectively, which would contribute only 0.5\% to the total luminosity of the quadruple system. When the LTT periods are 
very short, the eclipse timing variations could be caused by the geometrical and perturbative effects of the circumbinary companions 
(Borkovits et al. 2011, 2013; Rappaport et al. 2013). We computed the semi-amplitudes of the dynamic perturbations (physical delays) 
on the motion of the eclipsing pair to be 0.00005 d and 0.000002 d, respectively, for the third and fourth components and found 
that their contributions are not significant. 

Alternatively, the periodic oscillations in the eclipse timing diagram might be caused by a magnetic activity cycle for 
systems with a spectral type later than F5 (Applegate 1992; Lanza et al. 1998). However, it is difficult for the Applegate model 
to produce a perfectly smooth periodic component in the timing residuals. Moreover, Lanza (2006) presented that this mechanism 
is not adequate in explaining the period modulation of close binaries with a late-type secondary. Thus, the periodic variations 
most likely arise from the LTT effects due to the existence of the two circumbinary companions that are gravitationally bound 
to the close eclipsing pair. These indicate that V404 Lyr is a quadruple system.

The positive coefficient of the quadratic term ($A$) listed in Table 2 indicates a secular period increase with a rate of 
$+$1.41 $\times 10^{-7}$ d yr$^{-1}$. The most common explanation of the parabolic variation is a mass transfer from 
the less massive secondary to the primary component if the secondary fills its inner Roche lobe. Our light-curve synthesis 
shows that the eclipsing pair of V404 Lyr is a semi-detached binary with a secondary-filling configuration (cf. Section 3). 
This indicates that the secondary to primary mass transfer can be possible in the system. Under the assumption of 
a conservative mass transfer, the transfer rate is 5.42$\times$10$^{-8}$ M$_\odot$ yr$^{-1}$. The observed value is approximately
40\% smaller when compared with the predicted rate of 9.06$\times$10$^{-8}$ M$_\odot$ yr$^{-1}$, which is calculated by 
assuming that the secondary transfers its present mass to the primary in a thermal time scale. Thus, the possible explanation 
of the secular variation might be some combination of non-conservative mass transfer and angular momentum loss due to 
magnetic braking.

\section{LIGHT-CURVE SYNTHESIS AND ABSOLUTE DIMENSIONS}

In Figure 3, the {\it Kepler} light curve of V404 Lyr is plotted as normalized flux {\it versus} orbital phase. Its shape 
resembles that of $\beta$ Lyr and therefore indicates a significant temperature difference between the two components and 
a significant distortion of the photospheres. In order to derive the physical parameters of this system, the {\it Kepler} data 
were analyzed using the 2007 version\footnote {ftp://ftp.astro.ufl.edu/pub/wilson/} of the Wilson-Devinney binary code 
(Wilson \& Devinney 1971, hereafter W-D). Independently of eclipse timing diagrams, this version can extract 
third-body orbit parameters ($a^{\prime}$, $i^{\prime}$, $e^{\prime}$, $\omega^{\prime}$, $P^{\prime}$, $T_{\rm c}^{\prime}$)
that revolve around an eclipsing pair from whole light curves (van Hamme \& Wlson 2007), where the primed notation refers 
to the third-body orbit in the W-D code. The parameters $i^{\prime}$, $e^{\prime}$, $\omega^{\prime}$ are the inclination, 
eccentricity, and argument of periastron of the third-body orbit, respectively. $a^{\prime}$=$a_{12}+a_{3}$ is 
the semimajor axis of the outer relative orbit, and $T_{\rm c}^{\prime}$ is the time of the superior conjunction of 
the mass center of the close binary with respect to the third body.

For the light-curve synthesis, the effective temperature of the hotter and presumably more massive primary star was 
initialized at $T_{1}$=6,561 K from the {\it Kepler} Input Catalogue (Brown et al. 2011). The logarithmic bolometric ($X$, $Y$) 
and monochromatic ($x$, $y$) limb-darkening coefficient were interpolated from the values of van Hamme (1993) in concert with 
the model atmosphere option. The gravity-darkening exponents and bolometric albedoes were fixed at standard values 
($g$=0.32 and $A$=0.5) for stars with convective envelopes, as surmised from the components' temperatures. Furthermore, 
a synchronous rotation for both components and a circular orbit were adopted, and the detailed reflection effect and 
third light source ($l_3$) were considered throughout the analyses. The values with parenthesized errors in Table 3 
signify adjusted parameters. In this paper, subscripts 1 and 2 refer to the primary and secondary stars eclipsed at Min I 
(at phase 0.0) and Min II, respectively.

Although the photometric studies of V404 Lyr were performed by Somosv\'ari \& Csizmadia (2006) and Pr\v sa et al. (2011), 
their results are preliminary and there is still no spectroscopic mass ratio ($q$). In order to generally understand  
the geometrical structure and photometric parameters of the system, an extensive $q$-search procedure was conducted for 
various modes of the W-D code, without permitting third body or spot effects. In this procedure, we considered 
the orbital inclination ($i$), effective temperature ($T$), dimensionless surface potential ($\Omega$), and luminosity ($L_1$). 
Because the {\it Kepler} data base is very large, we used 1,000 normal points that were calculated using bin widths of 0.001 
in phase from all individual observations and were assigned weights equal to the number of observations per normal point. 
The $q$-search result exhibited acceptable photometric solutions for mode 5 only, which are semi-detached systems 
with the secondary filling its inner Roche lobe. As seen in Figure 4, the weighted sum of the squared residuals 
($\Sigma W(O-C)^2$) reached a minimum value at $q$=0.38. For subsequent calculations, all individual points were used and 
the mass ratio and third-body parameters were considered to be additional free variables. The results are listed in the second 
and third columns of Table 3 and they appear as a dashed curve in the top panel of Figure 3. The light residuals from 
the solution are calculated to see the details of non-modeled lights and plotted in the middle panel of the figure, wherein 
it can be seen that the unspotted model does not describe the observed light curve around phases 0.39 and 0.61. 

In short-period semi-detached binaries, the light discrepancy may result from either a mass transfer between the component stars 
or a magnetic dynamo in the systems with deep convective envelopes. Therefore, we tested three possible spot models: 
a hot spot on the primary star due to the mass transfer and a cool spot on either binary component caused by magnetic activity. 
The best solution is given in the fourth and fifth columns of Table 3 and is described using a solid curve in the top panel of 
Figure 3. The residuals from this spot model are plotted in the bottom panel of the figure. Separate trials for other spot 
configurations were not as successful as the fitting provided in Table 3. From the table and displays, it is clearly seen that 
the photometric parameters without and with a spot are very close to each other and that the cool spot on the secondary component 
could almost entirely explain the light variation. The third-body orbit parameters computed from all {\it Kepler} data are 
in good agreement with those from the historical eclipse timings within the uncertainties. In all procedures, we searched for 
a possible third light but the parameter remained indistinguishable from zero within its error. As discussed in the previous section, 
the light contribution of the third body to the multiple system is approximately 0.5 \%, assuming the binary and 
third-body coplanar orbit; thus, the absence of this evidence is not surprising.

The differential correction (DC) program of the W-D code produces the errors computed from the covariance matrix using 
the standard method. The parameter errors should arise from a final run that includes all adjustable parameters together. However, 
some papers (e.g., Maceroni \& Rucinski 1997) commented that the error estimates in the output from DC are unrealistically 
small due to the strong correlations between the relatively numerous parameters and the non-normal distribution of the measurement 
errors. Following the procedure described by Koo et al. (2014), we estimated the errors for the fitted parameters. For this purpose, 
the {\it Kepler} data were split into ten subsets and modeled individually with the W-D code. Then, the error estimates 
were calculated as the standard deviations of the ten different values for each parameter. The errors in Table 3 are 
the 1$\sigma$-values adopted from the larger of the values from the ten datasets and from all data.

The absolute dimensions of V404 Lyr can be estimated from our photometric solutions and empirical relations between spectral type
and mass. Assuming that the primary component is a normal main-sequence star with a spectral type of F4 V from its temperature 
(Harmanec 1988), we obtained the absolute parameters for the eclipsing pair given in the bottom of Table 3. 
The luminosity ($L$) and bolometric magnitudes ($M_{\rm bol}$) were computed by adopting $T_{\rm eff}$$_\odot$=5,780 K and 
$M_{\rm bol}$$_\odot$=+4.73 as solar values. For the absolute visual magnitudes ($M_{\rm V}$), we used the bolometric 
corrections (BCs) appropriate for the temperature of each component from the relationship between $\log T$ and BC given by 
Torres (2010). With an apparent magnitude of $V$=+11.39 and an interstellar absorption of $A_{\rm V}$=0.30 
(Csizmadia \& S\'andor 2001), we calculated a nominal distance to the system of 465 pc. This is too small compared with 
the distance (4,188 pc) of NGC 6791 (Basu et al. 2011) and, hence, V404 Lyr is not a member of the open cluster.

\section{LIGHT RESIDUALS AND PULSATIONAL CHARACTERISTICS}

Figure 5 plots the light curve residuals from our cool-spot model distributed in BJD instead of orbital phase as in Figure 3, 
wherein the lower panel presents a short section of the residuals. This figure clearly exhibits light variations with a semi-amplitude 
of about 30 mmag maximum. The amplitude changes from cycle to cycle, which indicates that multiple curves with different periods 
are superimposed. We performed multiple frequency analyses for the residuals using the PERIOD04 program (Lenz \& Breger 2005). 
The binary components block each other's lights during eclipses, which can make complicated variations. In order to remove 
this unmanageable effect, we analyzed only out-of-eclipse residuals (phases 0.134$-$0.366 and 0.634$-$0.866) after removing 
the times of the primary and secondary eclipses. The top panel of Figure 6 displayed the amplitude spectra in the frequency range 
from 0 to the Nyquist limit of 24.47 d$^{-1}$. Because only the out-of-eclipse phase data was used, the orbital frequency 
$f_{\rm orb}$ caused alias effects, i.e. 1.368 d$^{-1}$ multiplets deviated from the dominant peak, which is seen in the figure. 
The main signals lie in the frequency region between 0.5 and 5.0 d$^{-1}$. 

After consecutive pre-whitening processes, a total of 40 frequencies were detected based on the criterion of the signal to 
noise amplitude ratio (S/N) larger than 4.0 (Breger et al. 1993). At each pre-whitening procedure, we applied a multiperiodic 
least-squares fit to the light residuals using the equation of $Z$ = $Z_0$ + $\Sigma _{i}$ $A_i \sin$(2$\pi f_i t + \phi _i$). 
Here, $Z$ and $Z_0$ denote the calculated magnitude and zero point, $A_i$ and $\phi _i$ are the amplitude and phase of 
the $i$th frequency, and $t$ is the time of each measurement. The amplitude spectra after pre-whitening the first six frequencies 
and then all 40 frequencies are presented in the middle and bottom panels of Figure 6, respectively. The results are listed in Table 4.
The uncertainties in the table were derived according to Kallinger et al. (2008). The synthetic curve computed from 
the 40-frequency fit is displayed in Figure 5.

We examined the frequency variations with time through analyzing the out-of-eclipse residuals at intervals of approximately 100 d. 
The 12 subsets resulted in a little differences from each other and 9$-$16 frequencies were detected at each subset with the same 
criterion of S/N$>$4.0. Figure 7 displayed the variability of the 12 frequencies detected most repeatedly. The eight frequencies 
of $f_1$$-$$f_6$, $f_{13}$, and $f_{17}$ are almost constant. On the contrary, the four frequencies of $f_7$$-$$f_9$ and $f_{18}$ 
are varied significantly, which might be partly affected by the changes in the spot parameters with time and/or by the irregularity 
in the mass transfer from the secondary component to the primary. These unequable phenomena may cause several other frequencies 
near the four frequencies that were detected in the full out-of-eclipse data in Table 4.

In order to determine which component caused the seven dominant and constant frequencies near 2.0 d$^{-1}$, we compared 
the light variations between the out-of-eclipse and primary/secondary eclipse data. A section of the light residuals from Figure 5 
is magnified in Figure 8, where the synthetic light curve calculated from the parameters in Table 4 is well matched with both 
the out-of-eclipse and secondary eclipse data. However, the curve is poorly fitted with the primary eclipse data. In our light-curve 
representation for V404 Lyr, the primary eclipse is a partial transit, while the secondary eclipse is a total occultation of 
the secondary star by the primary. Therefore, the agreement in the secondary eclipse but disagreement in the primary eclipse 
implies that the primary star is the main light variation source. This discrepancy can be explained by the phase shift of 
non-radial pulsation during the eclipse as simulated by Reed et al. (2005).

Considering the physical parameters in Table 3, the primary component of V404 Lyr is an early F-type main sequence star,
located near the red edge of instability strip for $\gamma$ Dor type pulsating stars (Handler \& Shobbrook 2002). 
The multiperiodic pulsations with low amplitudes less than 10 mmag can occur in this type of pulsators. We calculated 
the pulsation constants of seven frequencies near 2.0 d$^{-1}$, $f_1$$-$$f_6$ and  $f_{13}$, from the equation of 
$\log Q_i = -\log f_i + 0.5 \log g + 0.1M_{\rm bol} + \log T_{\rm eff} - 6.456$ (Breger 2000). The results are listed in 
the third column of Table 5. The $Q$s of 0.24$-$0.27 d correspond to the gravity modes of $\gamma$ Dor pulsating stars.
The ratios of the orbital frequency to the seven pulsation frequencies $f_{\rm orb}$:$f_{\rm 1-6,13}$ are very close to 2:3,
which implies that the $\gamma$ Dor type pulsations might be excited by the tidal interaction of the close secondary component.
 
We attempted to identify the pulsation modes of the seven frequencies by applying the Frequency Ratio Method (FRM), which 
was introduced by Moya et al. (2005). According to the asymptotic approximation (Tassoul 1980) assuming adiabatic and 
non-rotation condition, the angular frequency of high-order ($n$$\gg$1) low-degree ($\ell$$\le$3) gravity modes for 
typical $\gamma$ Dor stars can be simplified as follows:
\begin{eqnarray}
\sigma_{n,\ell} \approx {\sqrt{\ell(\ell+1)} \over (n+1/2) \pi} {\cal J} ,
\end{eqnarray}
where $\cal J$ is the integral of the Brunt-V\"ais\"al\"a frequency. The theoretical value of $\cal J$ is approximated as a constant 
for gravity modes of typical $\gamma$ Dor stars (Moya et al. 2005). Therefore, the frequency ratio is expressed as follows:
\begin{eqnarray}
{\sigma_{n_1,\ell_1} \over \sigma_{n_2,\ell_2}} \approx {(n_2+1/2) \over (n_1+1/2)} {\sqrt{\ell_1(\ell_1+1)} \over \sqrt{\ell_2(\ell_2+1)}} .
\end{eqnarray}
We calculated the model frequency ratios from equation (4) for several pulsation mode sets ($n$, $\ell$) and found out 
the best fitted ratios with the observed ratios, as listed at Table 5. Although the rotation velocity of V404 Lyr has not been 
known, the primary component may be a fast rotator of approximately 120 km s$^{-1}$ assuming a synchronized rotation with 
the orbital motion. Because the FRM intrinsic error increases with rotational velocity (Su\'arez et al. 2005), the ratio error 
was set to have a slightly larger value of $\pm$0.02; Rodriguez et al. (2006) set the error to be $\pm$0.012 for HD 218427 with 
a moderate rotation velocity of 72 km s$^{-1}$. The four frequencies of $f_2$, $f_5$, $f_1$, and $f_4$ are identified as 
$\ell$=2 mode for four consecutive radial orders of $n$=24, 25, 26, and 27, respectively. And the two frequencies $f_3$ and $f_6$ 
are $\ell$=1 mode for two orders of $n$=14 and 15, respectively. The other $f_{13}$ with a lower amplitude than the former six 
frequencies may be excited by a less-detectable $\ell$=3 mode for $n$=39.

We obtained the observed frequency integral $\cal J_{\rm obs}$=771.5 $\pm$ 4.2 $\mu$Hz from equation (3), averaging 
the values for the seven frequencies. This is similar to the theoretical value $\cal J_{\rm theo}$ near 750 $\mu$Hz for a model 
of $\log$ $T_{\rm eff}$=3.817, 1.35 $M_\odot$, and [Fe/H]=0.0 (see Figure 5 in the paper by Moya et al. 2005). 
Considering that the physical properties of V404 Lyr, such as metal abundance and effective temperature, are uncertain, 
the difference of about 20 $\mu$Hz may be acceptable.

\section{DISCUSSION AND CONCLUSIONS}

We have presented the long-term photometric behavior of V404 Lyr which demonstrates total eclipses and multiperiodic pulsations, 
based on all available data as well as the ${\it Kepler}$ data. Our analyses of 2,922 eclipse timings, spanning more than 18 yrs, 
reveal that the orbital period of the system has varied due to two periodic variations superimposed on an upward-opening parabola,
with cycle lengths of $P_3$=649 d and $P_4$=2,154 d and semi-amplitudes of $K_3$=193 s and $K_4$=49 s, respectively. 
The upward parabola indicates a continuous period increase, which can be plausibly explained by a combination of mass transfer 
from the secondary to the primary star and angular momentum loss. The most reasonable explanation for both cycles is a pair of 
LTT effects driven by the presence of a third and fourth component with minimum masses of $M_3$=0.47 M$_\odot$ and 
$M_4$=0.047 M$_\odot$. As long as the inclination of the orbital plane of the fourth component is higher than 43$^\circ$, the mass 
of this object is below the theoretical threshold of $\sim$0.07 M$_\odot$ for a hydrogen-burning star. 

In eclipse timing diagrams, it is possible that the periodic variations with small amplitudes may be produced by 
the sporadic asymmetries of eclipse light curves due to stellar activity such as starspots and pulsations (cf. Lee et al. 2014). 
As in the process described by Lee et al. (2013b), we combined the {\it Kepler} data at intervals of 10 orbital periods 
($\sim$ 7.3 d) and calculated 174 minimum epochs for those datasets using the W-D code through adjusting only 
the ephemeris epoch ($T_0$) in the spot model of Table 3. The results are illustrated with the `x' symbols in Figure 2. It can 
be seen that the light-curve timings calculated from the W-D code agree well with our analysis of the eclipse timing variation. 
Thus, we conclude that the orbital period of V404 Lyr has varied due to two periodic oscillations plus an upward parabola.

The ${\it Kepler}$ light curves of V404 Lyr were satisfactorily modeled using a cool spot on the secondary component and 
third-body parameters. The results represent the eclipsing pair as a semi-detached binary with $q$=0.382, $i$=83$^\circ$, and 
$\Delta T$=1,193 K, in which the primary component fills its limiting lobe by approximately 93\%. The cool spot may be produced 
by magnetic dynamo-related activity because the system is rotating rapidly and the components have a deep convective envelope. 
The third-body parameters from the W-D code are consistent with those calculated from the eclipse timings themselves. 
From the absolute parameters given in Section 3, it is possible to estimate the evolutionary state of the binary system
in the mass-radius, mass-luminosity, and Hertzsprung-Russell (HR) diagrams (Hilditch et al. 1988; \. Ibano\v{g}lu et al. 2006).
In these diagrams, the primary star lies in the main-sequence band, while the secondary is clearly beyond the terminal-age 
main sequence and its radius is more than two times oversized compared with the main-sequence stars of the same mass. Furthermore, 
the locations of the two component stars fall amid those of previously-known near-contact binaries (Shaw 1990, 1994). Thus, 
the eclipsing pair is a semi-detached and FO Vir-type near-contact binary consisting of a detached main-sequence primary star 
and an evolved lobe-filling secondary component. 

Through analyzing the light residuals from our W-D binary model, we found intrinsic variations of 40 frequencies between 0.60 and 
4.11 d$^{-1}$ with amplitudes from 0.49 to 5.72 mmag. Among these, seven frequencies near 2.0 d$^{-1}$ with large amplitudes 
and high stability for about 4 years transpired to originate from the pulsation of the primary component. $\delta$ Sct pulsators 
are stars pulsating in low-order pressure modes with short periods of 0.02$-$0.2 d (Breger 2000), whereas $\gamma$ Dor pulsators 
are A-F stars of luminosity class IV-V pulsating in high-order gravity modes with typical periods in the range 0.4$-$3 d 
(Kaye et al. 1999, Henry et al. 2005). The pulsation periods of V404 Lyr indicate that the primary component is a candidate 
for $\gamma$ Dor type pulsating stars, rather than $\delta$ Sct variables. Handler \& Shobbrook (2002) argued the relationship 
between the $\delta$ Sct and $\gamma$ Dor variables and showed that the two pulsators are clearly separated by 
their pulsation constants: the $\delta$ Sct stars have $Q <$ 0.04 d and the $\gamma$ Dor stars $Q >$ 0.23 d. As listed in 
the third column of Table 5, the pulsation constants of V404 Lyr are in the range of 0.24$-$0.27 d. A recent study by 
Zhang et al. (2013) indicates that the ratio of pulsation and orbital periods could function as a criterion to distinguish them 
in EBs, where $P_{\rm pul}/P_{\rm orb}=$0.09 is the upper limit for $\delta$ Sct stars. All of these results reveal that 
the primary component is a $\gamma$ Dor pulsator.

The eclipsing pair of V404 Lyr is a short-period semi-detached binary and its pulsating component has the largest filling factor
(93\%) of the 78 known EBs with $\delta$ Sct or $\gamma$ Dor pulsators (Zhang et al. 2013; Maceroni et al. 2014; Yang et al. 2014). 
Based on this, the pulsating characteristics of the primary star may be partly influenced by tidal interaction and secondary 
to primary mass transfer. The properties of a convective core in the $\gamma$ Dor star can be demonstrated through studying 
tidally induced and gravity mode pulsations. Because it has evolved with mass accretion between the binary components, 
the pulsating primary star may have a different evolutionary history than single $\gamma$ Dor pulsators. 
High-resolution spectroscopy will assist in determining the absolute parameters of the multiple system and in understanding 
its evolutionary status better than is possible with photometry alone.

\acknowledgments{ }
This research has made use of the {\it Kepler} and SuperWASP public archives. {\it Kepler} was selected as the 10th mission of 
the Discovery Program. Funding for the {\it Kepler} mission is provided by the NASA Science Mission directorate. 
The WASP consortium comprises of the University of Cambridge, Keele University, University of Leicester, The Open University, 
The Queen's University Belfast, St. Andrews University and the Isaac Newton Group. Funding for WASP comes from the consortium 
universities and from the UK's Science and Technology Facilities Council. We have used the Simbad database maintained at CDS, 
Strasbourg, France. This work was supported by the KASI (Korea Astronomy and Space Science Institute) grant 2014-1-400-06.

\clearpage
\begin{figure}
\includegraphics[]{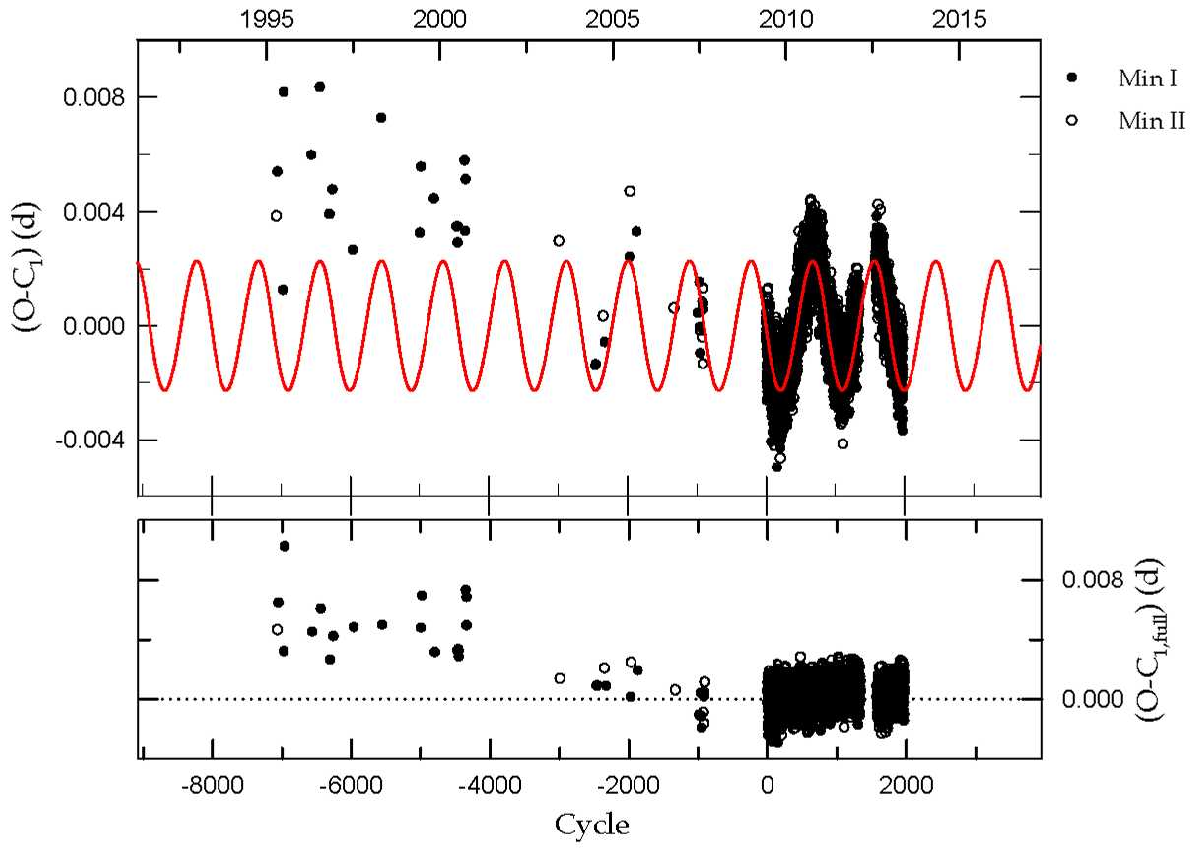}
\caption{$O$--$C$ diagram of V404 Lyr. In the upper panel, constructed with the linear terms of equation (1), the continuous curve 
represents the LTT orbit. The residuals from this ephemeris are plotted in the lower panel. }
\label{Fig1}
\end{figure}

\begin{figure}
\includegraphics[]{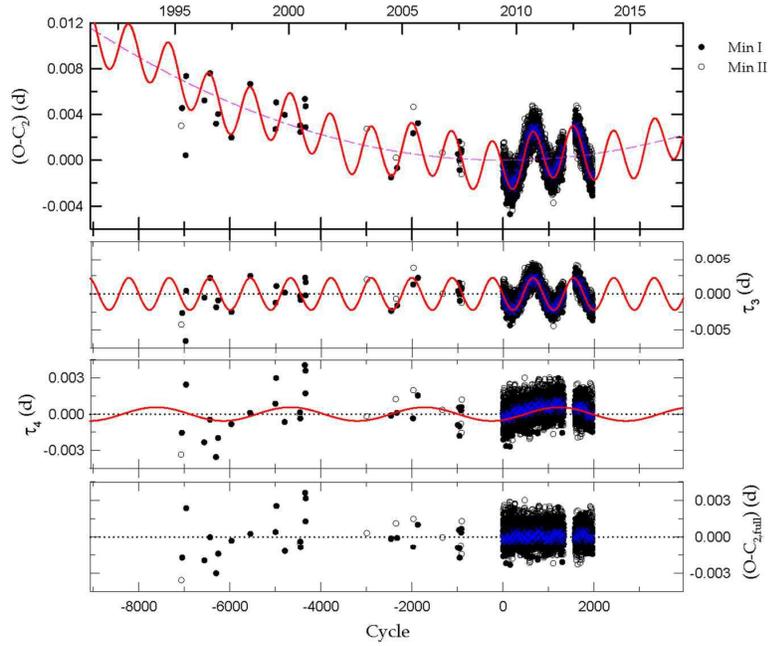}
\caption{In the top panel the $O$--$C$ diagram of V404 Lyr with respect to the linear terms of the quadratic {\it plus} 
two-LTT ephemeris. The full ephemeris is drawn as a solid curve and the dashed parabola results from the quadratic term 
of equation (2). The second and third panels display the short- and long-term LTT orbits, respectively. The bottom panel 
represents the residuals from the complete ephemeris. In all panels, the blue `x' symbols refer to the minimum times 
obtained by means of analyzing 174 {\it Kepler} datasets with the W-D code. }
\label{Fig2}
\end{figure}

\begin{figure}
\includegraphics[]{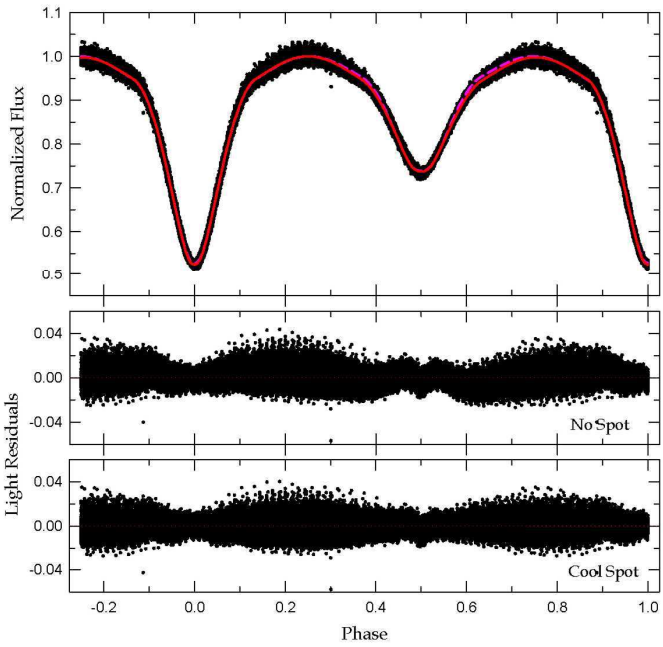}
\caption{Light curve of V404 Lyr with the fitted models. The circles are individual measurements from the {\it Kepler} spacecraft 
and the dashed and solid lines represent the synthetic curves obtained from no spot and the cool-spot model on the secondary star,
respectively. The light residuals corresponding to the two models are plotted in the middle and bottom panels. }
\label{Fig3}
\end{figure}

\begin{figure}
\includegraphics[]{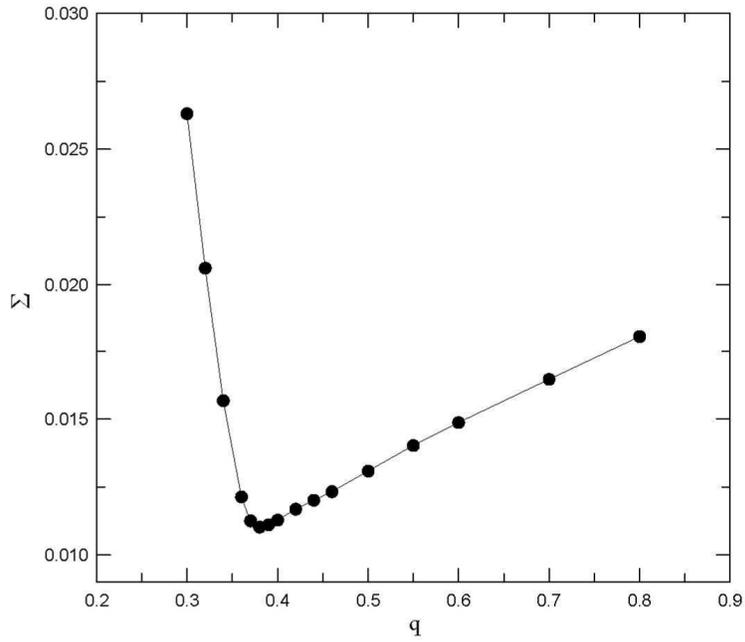}
\caption{Behavior of $\Sigma$ (the sum of the residuals squared) of V404 Lyr as a function of mass ratio $q$, showing 
a minimum value at $q$=0.38. }
\label{Fig4}
\end{figure}

\begin{figure}
\includegraphics[]{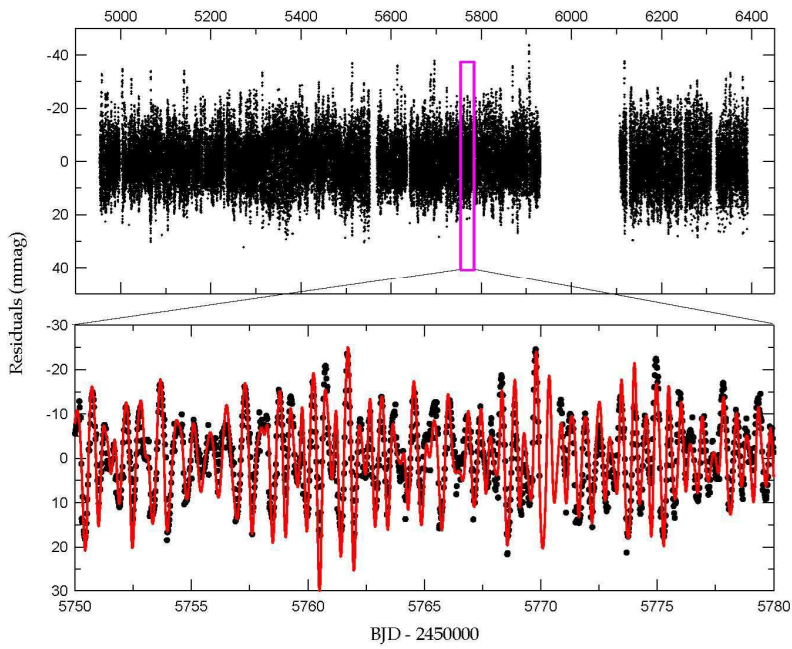}
\caption{Light curve residuals distributed in BJD instead of the phase, as in Figure 3. The lower panel presents a short section of 
the residuals marked using the inset box of the upper panel. The synthetic curve was computed from the 40-frequency fit to the data. }
\label{Fig5}
\end{figure}

\begin{figure}
\includegraphics[]{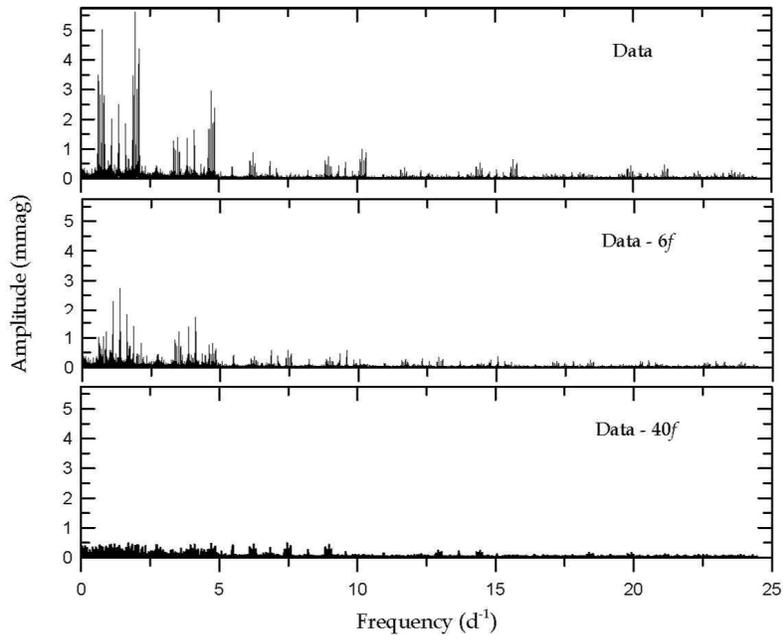}
\caption{Amplitude spectra before (top panel) and after pre-whitening the first six frequencies (middle) and all 40 frequencies 
(bottom) from the PERIOD04 program for the out-of-eclipse residual data. }
\label{Fig6}
\end{figure}

\begin{figure}
\includegraphics[]{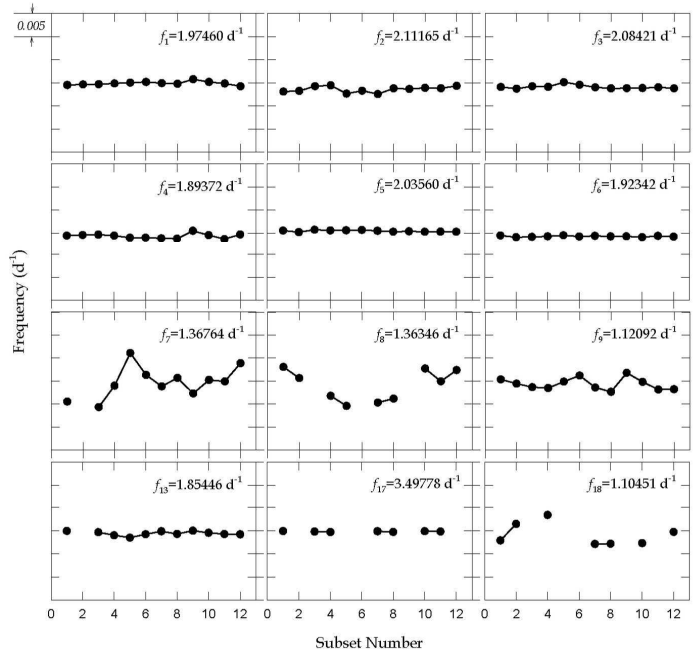}
\caption{Variability of the main frequencies detected in the 12 subsets at intervals of $\sim$100 d. In all panels, the y-axes are 
scaled as 0.03 d$^{-1}$ and the tick intervals are 0.005 d$^{-1}$. }
\label{Fig7}
\end{figure}

\begin{figure}
\includegraphics[]{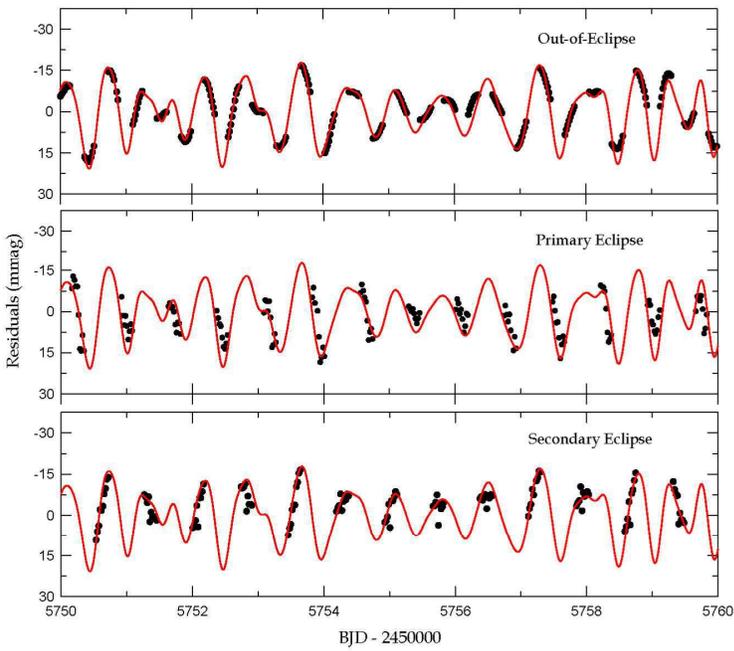}
\caption{Amplified images of a section of the light residuals in Figure 5, divided into the times of the outside, primary, 
and secondary eclipses. }
\label{Fig8}
\end{figure}

\clearpage
\begin{deluxetable}{lcrccc}
\tablewidth{0pt}
\tablecaption{SuperWASP and {\it Kepler} Eclipse Timings for V404 Lyr}
\tablehead{
\colhead{BJD}    & \colhead{Error} & \colhead{Epoch} & \colhead{$O$--$C_{1,\rm full}$} & \colhead{$O$--$C_{2,\rm full}$} & \colhead{Min}
}
\startdata
2,454,230.63111  & $\pm$0.00031    & $-$989.0     & $-$0.00103  & $-$0.00092  & I       \\
2,454,249.63673  & $\pm$0.00077    & $-$963.0     & $+$0.00042  & $+$0.00057  & I       \\
2,454,252.55891  & $\pm$0.00046    & $-$959.0     & $-$0.00110  & $-$0.00096  & I       \\
2,454,256.57899  & $\pm$0.00070    & $-$953.5     & $-$0.00113  & $-$0.00098  & II      \\
2,454,257.67461  & $\pm$0.00038    & $-$952.0     & $-$0.00190  & $-$0.00176  & I       \\
2,454,278.50706  & $\pm$0.00077    & $-$923.5     & $-$0.00087  & $-$0.00072  & II      \\
2,454,279.60470  & $\pm$0.00048    & $-$922.0     & $+$0.00037  & $+$0.00053  & I       \\
2,454,282.52827  & $\pm$0.00044    & $-$918.0     & $+$0.00024  & $+$0.00039  & I       \\
2,454,283.62463  & $\pm$0.00049    & $-$916.5     & $+$0.00021  & $+$0.00036  & II      \\
2,454,286.54651  & $\pm$0.00044    & $-$912.5     & $-$0.00161  & $-$0.00146  & II      \\
2,454,287.64502  & $\pm$0.00032    & $-$911.0     & $+$0.00050  & $+$0.00066  & I       \\
2,454,291.66576  & $\pm$0.00050    & $-$905.5     & $+$0.00115  & $+$0.00130  & II      \\
2,454,953.89883  & $\pm$0.00030    &   $+$0.5     & $-$0.00020  & $+$0.00057  & II      \\ 
2,454,954.26383  & $\pm$0.00022    &   $+$1.0     & $-$0.00066  & $+$0.00011  & I       \\ 
2,454,954.63106  & $\pm$0.00079    &   $+$1.5     & $+$0.00109  & $+$0.00187  & II      \\ 
2,454,954.99363  & $\pm$0.00070    &   $+$2.0     & $-$0.00179  & $-$0.00102  & I       \\ 
2,454,955.36131  & $\pm$0.00042    &   $+$2.5     & $+$0.00042  & $+$0.00119  & II      \\ 
2,454,955.72618  & $\pm$0.00029    &   $+$3.0     & $-$0.00017  & $+$0.00060  & I       \\ 
2,454,956.09245  & $\pm$0.00070    &   $+$3.5     & $+$0.00063  & $+$0.00140  & II      \\ 
2,454,956.45536  & $\pm$0.00073    &   $+$4.0     & $-$0.00191  & $-$0.00115  & I       \\ 
\enddata
\tablecomments{This table is available in its entirety in machine-readable and Virtual Observatory (VO) forms in the online journal. 
A portion is shown here for guidance regarding its form and content.}
\end{deluxetable}

\begin{deluxetable}{lccccc}
\tablewidth{0pt}
\tablecaption{Parameters for the LTT Orbits of V404 Lyr}
\tablehead{
\colhead{Parameter}      &  \colhead{Single-LTT}  &&  \multicolumn{2}{c}{Quadratic {\it plus} Two-LTT}     &  \colhead{Unit}        \\ [1.5mm] \cline{4-5} \\ [-2.0ex]
\colhead{}               &  \colhead{$\tau_{3}$}  &&  \colhead{$\tau_{3}$}      & \colhead{$\tau_{4}$}     &                         
}                                                                                                                                    
\startdata                                                                                                                           
$T_0$                    &  2,454,953.53370(24)   &&  \multicolumn{2}{c}{2,454,953.53346(12)}              &   BJD                  \\
$P$                      &  0.73094341(21)        &&  \multicolumn{2}{c}{0.73094326(11)}                   &   d                    \\
$a_{12}\sin i_{3,4}$     &  0.395(57)             &&  0.387(30)                 &  0.098(31)               &   au                   \\
$\omega$                 &  153.1(8.7)            &&  43.6(4.5)                 &  191(17)                 &   deg                  \\
$e$                      &  0.08(27)              &&  0.05(11)                  &  0.00(54)                &                        \\
$n$                      &  0.5543(98)            &&  0.5549(44)                &  0.167(12)               &   deg d$^{-1}$         \\
$T_{\rm peri}$           &  2,454,903(15)         &&  2,454,703(8)              &  2,454,301(101)          &   BJD                  \\
$P_{3,4}$                &  649(11)               &&  648.8(5.1)                &  2154(160)               &   d                    \\
$K$                      &  197(28)               &&  193(15)                   &  49(16)                  &   s                    \\
$f(M_{3,4})$             &  0.0195(28)            &&  0.0184(14)                &  0.0000272(89)           &   M$_\odot$            \\
$M_{3,4} \sin i_{3,4}$   &  0.479(38)             &&  0.469(20)                 &  0.0468(75)              &   M$_\odot$            \\
$a_{3,4} \sin i_{3,4}$   &  1.541(62)             &&  1.545(32)                 &  4.91(42)                &   au                   \\[0.5mm]
$A$                      &  \dots                 && \multicolumn{2}{c}{$+1.41(49)\times 10^{-10}$}        &   d                    \\
$dP$/$dt$                &  \dots                 && \multicolumn{2}{c}{$+1.41(49)\times 10^{-7}$}         &   d yr$^{-1}$          \\[0.5mm]
Reduced $\chi^2$         &  2.256                 &&  \multicolumn{2}{c}{0.996}                            &                        \\
\enddata
\tablecomments{The parameters $P_{3,4}$ and $K$ are the LTT periods and semi-amplitudes, and $f(M_{3,4})$, $M_{3,4} \sin i_{3,4}$, 
and $a_{3,4} \sin i_{3,4}$ are the mass functions, masses, and semi-major axes of the circumbinary objects. }
\end{deluxetable}

\begin{deluxetable}{lccccc}
\tabletypesize{\scriptsize} %{\small}
\tablewidth{0pt} 
\tablecaption{Physical Parameters of V404 Lyr}
\tablehead{
\colhead{Parameter}                      & \multicolumn{2}{c}{Without Spot}            && \multicolumn{2}{c}{With Spot}               \\ [1.0mm] \cline{2-3} \cline{5-6} \\[-2.0ex]
                                         & \colhead{Primary} & \colhead{Secondary}     && \colhead{Primary} & \colhead{Secondary}         
}
\startdata 
$T_0$ (BJD)                              & \multicolumn{2}{c}{2,454,953.53320(6)}      && \multicolumn{2}{c}{2,454,953.53308(6)}      \\
$P$ (d)                                  & \multicolumn{2}{c}{0.73094404(2)}           && \multicolumn{2}{c}{0.73094404(2)}           \\
$q$                                      & \multicolumn{2}{c}{0.3755(8)}               && \multicolumn{2}{c}{0.3815(6)}               \\
$i$ (deg)                                & \multicolumn{2}{c}{82.92(3)}                && \multicolumn{2}{c}{82.95(2)}                \\
$T$ (K)                                  & 6,554(69)         & 5,363(47)               && 6,555(67)         & 5,362(46)               \\
$\Omega$                                 & 2.8406(6)         & 2.6277                  && 2.8549(6)         & 2.6400                  \\
$X$, $Y$                                 & 0.638, 0.243      & 0.645, 0.186            && 0.638, 0.243      & 0.645, 0.185            \\
$x$, $y$                                 & 0.613, 0.285      & 0.693, 0.219            && 0.613, 0.285      & 0.693, 0.219            \\
$L$/($L_{1}$+$L_{2}$)                    & 0.8094(2)         & 0.1906                  && 0.8072(2)         & 0.1928                  \\
$r$ (pole)                               & 0.4013(1)         & 0.2778(1)               && 0.3999(1)         & 0.2789(1)               \\
$r$ (point)                              & 0.4605(2)         & 0.4009(1)               && 0.4588(2)         & 0.4024(1)               \\
$r$ (side)                               & 0.4215(1)         & 0.2895(1)               && 0.4199(1)         & 0.2907(1)               \\
$r$ (back)                               & 0.4380(1)         & 0.3222(1)               && 0.4364(1)         & 0.3234(1)               \\
$r$ (volume)                             & 0.4207            & 0.2976                  && 0.4192            & 0.2989                  \\ [1.0mm]
\multicolumn{6}{l}{Third-body parameters:}                                                                                            \\        
$a^{\prime}$($R_\odot$)                  & \multicolumn{2}{c}{364.4(6)}                && \multicolumn{2}{c}{364.2(6)}                \\        
$i^{\prime}$ (deg)                       & \multicolumn{2}{c}{82.9}                    && \multicolumn{2}{c}{82.9}                    \\        
$e^{\prime}$                             & \multicolumn{2}{c}{0.11(2)}                 && \multicolumn{2}{c}{0.11(2)}                 \\        
$\omega^{\prime}$  (deg)                 & \multicolumn{2}{c}{7(8)}                    && \multicolumn{2}{c}{8(7)}                    \\        
$P^{\prime}$ (d)                         & \multicolumn{2}{c}{647(1)}                  && \multicolumn{2}{c}{647(1)}                  \\        
$T_{\rm c}^{\prime}$ (BJD)               & \multicolumn{2}{c}{2,455,089(3)}            && \multicolumn{2}{c}{2,455,088(3)}            \\        
\multicolumn{6}{l}{Spot parameters:}                                                                                                  \\        
Colatitude (deg)                         & \dots             & \dots                   && \dots             & 60.7(5)                 \\        
Longitude (deg)                          & \dots             & \dots                   && \dots             & 352.9(2)                \\        
Radius (deg)                             & \dots             & \dots                   && \dots             & 22.9(7)                 \\        
$T$$\rm _{spot}$/$T$$\rm _{local}$       & \dots             & \dots                   && \dots             & 0.872(9)                \\
$\Sigma W(O-C)^2$                        & \multicolumn{2}{c}{0.0018}                  && \multicolumn{2}{c}{0.0015}                  \\ [1.0mm]
\multicolumn{6}{l}{Absolute parameters:}                                                                                              \\            
$M$($M_\odot$)                           & 1.35              &  0.51                   && 1.35              &  0.52                   \\
$R$($R_\odot$)                           & 1.76              &  1.25                   && 1.76              &  1.26                   \\
$\log$ $g$ (cgs)                         & 4.08              &  3.95                   && 4.08              &  3.95                   \\
$\rho$ (g cm$^3)$                        & 0.35              &  0.37                   && 0.35              &  0.37                   \\
$L$($L_\odot$)                           & 5.14              &  1.15                   && 5.12              &  1.17                   \\
$M_{\rm bol}$ (mag)                      & 2.95              &  4.57                   && 2.96              &  4.56                   \\
BC (mag)                                 & $+$0.01           &  $-$0.18                && $+$0.01           &  $-$0.18                \\
$M_{\rm V}$ (mag)                        & 2.94              &  4.75                   && 2.95              &  4.74                   \\
Distance (pc)                            & \multicolumn{2}{c}{465}                     && \multicolumn{2}{c}{465}                     \\
\enddata
\tablenotetext{a}{Mean volume radius.}
\end{deluxetable}

\begin{deluxetable}{lccccc}
\tabletypesize{\scriptsize} %{\small}
\tablewidth{0pt}
\tablecaption{Multiple Frequency Analysis of V404 Lyr$\rm ^a$}
\tablehead{
             & \colhead{Frequency}   & \colhead{Amplitude} & \colhead{Phase}   & \colhead{S/N$\rm ^b$}  & \colhead{Remark}      \\
             & \colhead{(d$^{-1}$)}  & \colhead{(mmag)}    & \colhead{(rad)}   &              & 
}
\startdata
$f_{1}$      & 1.97460$\pm$0.00001   & 5.72$\pm$0.03       & 3.62$\pm$0.01     & 49.69        & $\gamma$ Dor                    \\
$f_{2}$      & 2.11165$\pm$0.00001   & 4.37$\pm$0.04       & 3.35$\pm$0.02     & 38.00        & $\gamma$ Dor                    \\
$f_{3}$      & 2.08421$\pm$0.00001   & 3.88$\pm$0.04       & 4.43$\pm$0.02     & 33.78        & $\gamma$ Dor                    \\
$f_{4}$      & 1.89372$\pm$0.00002   & 3.35$\pm$0.05       & 2.58$\pm$0.02     & 28.96        & $\gamma$ Dor                    \\
$f_{5}$      & 2.03560$\pm$0.00002   & 3.38$\pm$0.05       & 1.86$\pm$0.02     & 29.45        & $\gamma$ Dor                    \\
$f_{6}$      & 1.92342$\pm$0.00002   & 2.86$\pm$0.06       & 5.12$\pm$0.03     & 24.79        & $\gamma$ Dor                    \\
$f_{7}$      & 1.36764$\pm$0.00003   & 1.68$\pm$0.11       & 5.80$\pm$0.05     & 13.22        & $\sim$$f_{\rm obs}$             \\
$f_{8}$      & 1.36346$\pm$0.00003   & 1.96$\pm$0.10       & 5.82$\pm$0.04     & 15.46        &                                 \\
$f_{9}$      & 1.12092$\pm$0.00003   & 2.28$\pm$0.09       & 2.22$\pm$0.04     & 17.03        &                                 \\
$f_{10}$     & 1.36245$\pm$0.00002   & 2.29$\pm$0.08       & 0.96$\pm$0.04     & 18.09        &                                 \\
$f_{11}$     & 1.37661$\pm$0.00004   & 1.35$\pm$0.14       & 5.63$\pm$0.07     & 10.64        &                                 \\
$f_{12}$     & 1.36935$\pm$0.00005   & 1.15$\pm$0.16       & 5.65$\pm$0.08     &  9.05        &                                 \\
$f_{13}$     & 1.85446$\pm$0.00004   & 1.38$\pm$0.13       & 0.23$\pm$0.06     & 11.73        & $\gamma$ Dor                    \\
$f_{14}$     & 1.37938$\pm$0.00004   & 1.44$\pm$0.13       & 1.30$\pm$0.06     & 11.38        &                                 \\
$f_{15}$     & 1.37863$\pm$0.00004   & 1.48$\pm$0.13       & 1.17$\pm$0.06     & 11.71        &                                 \\
$f_{16}$     & 1.36011$\pm$0.00004   & 1.54$\pm$0.12       & 2.81$\pm$0.06     & 12.13        &                                 \\
$f_{17}$     & 3.49778$\pm$0.00004   & 1.17$\pm$0.12       & 1.91$\pm$0.06     & 11.99        & 4$f_{\rm orb} - f_1$            \\
$f_{18}$     & 1.10451$\pm$0.00005   & 1.13$\pm$0.18       & 1.84$\pm$0.08     &  8.39        &                                 \\
$f_{19}$     & 0.60658$\pm$0.00006   & 1.02$\pm$0.20       & 0.94$\pm$0.10     &  7.29        & $f_1- f_{\rm orb}$              \\
$f_{20}$     & 1.36522$\pm$0.00003   & 1.73$\pm$0.11       & 5.68$\pm$0.05     & 13.67        &                                 \\
$f_{21}$     & 3.36071$\pm$0.00005   & 0.94$\pm$0.16       & 2.90$\pm$0.07     &  9.40        & 4$f_{\rm orb} - f_2$            \\
$f_{22}$     & 1.38174$\pm$0.00006   & 1.02$\pm$0.18       & 2.76$\pm$0.09     &  8.10        &                                 \\
$f_{23}$     & 1.36591$\pm$0.00004   & 1.56$\pm$0.12       & 1.03$\pm$0.06     & 12.31        &                                 \\
$f_{24}$     & 3.38816$\pm$0.00006   & 0.80$\pm$0.18       & 4.53$\pm$0.09     &  8.13        & 4$f_{\rm orb} - f_3$            \\
$f_{25}$     & 3.43679$\pm$0.00006   & 0.74$\pm$0.20       & 0.19$\pm$0.09     &  7.49        & 4$f_{\rm orb} - f_5$            \\
$f_{26}$     & 1.70325$\pm$0.00008   & 0.72$\pm$0.25       & 0.60$\pm$0.12     &  5.85        & 2$f_4 - f_3$                    \\
$f_{27}$     & 1.12865$\pm$0.00009   & 0.69$\pm$0.29       & 1.38$\pm$0.13     &  5.22        &                                 \\
$f_{28}$     & 1.73292$\pm$0.00008   & 0.70$\pm$0.26       & 1.32$\pm$0.12     &  5.72        & $f_4 + f_6 - f_3$               \\
$f_{29}$     & 3.57865$\pm$0.00006   & 0.70$\pm$0.20       & 0.10$\pm$0.10     &  7.31        & 4$f_{\rm orb} - f_4$            \\
$f_{30}$     & 1.08647$\pm$0.00009   & 0.70$\pm$0.29       & 1.02$\pm$0.13     &  5.18        &                                 \\
$f_{31}$     & 1.38327$\pm$0.00008   & 0.73$\pm$0.26       & 4.17$\pm$0.12     &  5.80        &                                 \\
$f_{32}$     & 3.54895$\pm$0.00007   & 0.65$\pm$0.22       & 3.77$\pm$0.10     &  6.78        & 4$f_{\rm orb} - f_6$            \\
$f_{33}$     & 1.36649$\pm$0.00006   & 0.96$\pm$0.20       & 3.32$\pm$0.09     &  7.56        &                                 \\
$f_{34}$     & 1.09570$\pm$0.00010   & 0.63$\pm$0.32       & 0.75$\pm$0.15     &  4.70        &                                 \\
$f_{35}$     & 1.34995$\pm$0.00010   & 0.60$\pm$0.32       & 1.97$\pm$0.15     &  4.72        &                                 \\
$f_{36}$     & 4.10496$\pm$0.00006   & 0.61$\pm$0.20       & 4.48$\pm$0.09     &  7.39        & $\sim$3$f_{\rm obs}$            \\
$f_{37}$     & 1.86618$\pm$0.00010   & 0.55$\pm$0.32       & 0.94$\pm$0.15     &  4.72        & $f_3 + f_4 - f_2$               \\
$f_{38}$     & 1.37784$\pm$0.00010   & 0.59$\pm$0.32       & 0.75$\pm$0.15     &  4.67        &                                 \\
$f_{39}$     & 4.09674$\pm$0.00006   & 0.59$\pm$0.21       & 3.65$\pm$0.10     &  7.22        &                                 \\
$f_{40}$     & 1.99261$\pm$0.00011   & 0.49$\pm$0.35       & 5.44$\pm$0.16     &  4.22        & 3$f_{\rm orb} - f_2$            \\
\enddata                                                                                                                           
\tablenotetext{a}{Frequencies are listed in order of detection. }
\tablenotetext{b}{Calculated in a range of 5 d$^{-1}$ around each frequency. }
\end{deluxetable}

\newpage
\begin{deluxetable}{lccccccc}
\tabletypesize{\scriptsize} %{\small}
\tablewidth{0pt}
\tablecaption{$\gamma$ Dor-Type Pulsation Properties of V404 Lyr}
\tablehead{
       & \colhead{Frequency}   & \colhead{$Q$}  & \colhead{($f_i$/$f_2$)$_{\rm obs}$} & \colhead{mode ($n$, $\ell$)}  & \colhead{($f_i$/$f_2$)$_{\rm model}$} & \colhead{$\Delta$($f_i$/$f_2$)$_{\rm obs-model}$} & \colhead{$\cal J_{\rm obs}$}    \\
       & \colhead{(d$^{-1}$)}  & \colhead{(d)}  &                                     &                               &                                       &                                                   & \colhead{($\mu$Hz)}  
}
\startdata		
$f_1$    & 1.97460   & 0.25   & 0.9351   & (26, 2)    & 0.9245   & +0.0106              & 776.75            \\	
$f_2$    & 2.11165   & 0.24   & $-$      & (24, 2)    & $-$      & $-$                  & 767.97            \\		
$f_3$    & 2.08421   & 0.24   & 0.9870   & (14, 1)    & 0.9755   & +0.0115              & 777.01            \\	
$f_4$    & 1.89372   & 0.26   & 0.8968   & (27, 2)    & 0.8909   & +0.0059              & 773.05            \\	
$f_5$    & 2.03560   & 0.24   & 0.9640   & (25, 2)    & 0.9608   & +0.0032              & 770.53            \\
$f_6$    & 1.92342   & 0.26   & 0.9109   & (15, 1)    & 0.9126   & $-$0.0017            & 766.52            \\
$f_{13}$ & 1.85446   & 0.27   & 0.8782   & (39, 3)    & 0.8772   & +0.0010              & 768.88            \\
         &           &        &          &            & Average  & +0.0051$\pm$0.0053   & 771.53$\pm$4.19   \\  
\enddata                                                                                                                           
\end{deluxetable}

\end{document}